# Vision as Adaptive Epistemology

Ignazio Licata
ISEM, Institute for Scientific Methodology, Palermo, Italy
Ignazio.licata@ejtp.info

> *veritas est adaequatio rei et intellectus*
> Thomas Aquinas, *De Veritate*
>
> *The situation of complete certainty*
> *is reached only by observation of an*
> *infinite number of events (God's Eye)*
> Tibor Vámos

## 1. INTRODUCTION

In the last years the debate on complexity has been developing and developing in transdisciplinary way to meet the need of explanation for highly organized collective behaviors and sophisticated hierarchical arrangements in physical, biological, cognitive and social systems. Unfortunately, no clear definition has been reached, so complexity appears like an anti-reductionist paradigm in search of a theory.

In our short survey we aim to suggest a clarification in relation to the notions of computational and intrinsic emergence, and to show how the latter is deeply connected to the new Logical Openness Theory, an original extension of Gödel theorems to the model theory. The epistemological scenario we are going to make use of is that of the theory of vision, a particularly instructive one. Vision is an element of our primordial relationship with the world; consequently it comes as no surprise that carefully taking into consideration the processes of visual perception can lead us straight to some significant questions useful to delineate a natural history of knowledge. The common Greek etymological root of "theory" and "vision" sounds like a metaphor pointing out the analogy between the modalities of vision and those we use "to see and build the world" (N. Goodman, 1978), because them both can say us something about the central role of the observer and the semantic complexity of cognitive strategies.

## 2. REDUCTIONISM AND *NAÏVE* OBJECTIVISM

Most of the problems in focusing the notion of complexity just come from the unsuitable extension of that *naïve* objectivism which represents the conceptual driftage of reductionism. This one is very useful a tool which has guaranteed Physics a considerable success; but when it is regarded as the unique and universal method an hidden postulate, apparently innocuous and natural, comes out: the world is "out there", independent from the observer, organized by levels, explicable by means of a chain of theories logically connected and each description level can be derived from the previous one simply by using proper mathematical techniques and, at the most, "bridge-laws".

Such kind of complexity corresponds to the algorithmic complexity (Chaitin, 2007); it measures the information that a Turing machine has to process to solve a problem in relation to the processing time and space (program length). It is interesting noticing how Artificial Intelligence and the current Everything Theories share the same conception: a Laplace's demon (Hahn, 2005) can solve "mind" in purely syntactical terms just in the same way as it reduces the physical world variety to a nutshell of fundamental particles and interactions. Similarly, in such a Universe nothing authentically new can turn out, and the only detectable emergence is the computational one (detection of patterns) which is obtained by the fundamental algorithmic compression.

An example of the above-mentioned emergence is well represented by the non-linear chaotic systems. There, the long-rage predictability is missing, but it is possible the step-by-step computation of the system's trajectories in the phase space. Even if such ingenuous idea of reductionism has been radically criticized (Anderson, 1979; Laughlin, 2006 Laughlin e Pines, 1999; Laughlin, Pines et al., 2000) and it has been observed that this kind of description can only be fulfilled within classical systems (Licata, 2008a), *naïve* objectivism and the independence from the observer are still the "hidden" postulates in the scientific activity.

In order to find an alternative we have to look at a dynamic theory of relationship between the observer and the observed which takes into account the co-adaptive processes as well as the ecological nature of the mind/world relations.

## 3. VISION BETWEEN SYNTAX AND SEMANTICS

The difficulties in developing artificial vision devices have been strongly instructive in understanding the semantic features of the complexity involved in the process. As it is known, the symbolic-algorithmic approach of classical cognitivism has succeeded only in recognition of very simple shape and dynamics, while connectionism –thanks to bio-morph inspired parallel and distributed computing - has achieved larger success in recognition of even greatly complicated patterns. Anyway, there's a common problem with them both: the meaning of vision. To clarify this crucial point, it will be useful to shortly examine some of the salient outcomes in neurodynamics.

When a visual impulse related to an object (frequency, luminosity, dynamics and so on) hits the retina, the information is distributed on the receptor-fields of many specialized neural "agencies". The classical problem is to bring the activity of many cognitive agencies back to a perceptive act we experience as a single one. In these years, the coherence process which makes the neural agencies synchronized so as to respond in collective and unitary way to an object recognition –know as "feature binding" (Singer & Gray 1995; Varela et al., 1999) – has been experimentally verified. If we consider each neuron as a threshold non-linear device, it means that the coherence process needs the threshold rearrangement of each neuron so as to arrange its output according to the other neurons involved in the same perception act. S. Grossberg studied the general logic of such feed-back process by ART, Adaptive Resonance Theory (Grossberg, 1988; Levine, 2000). The input is triggered by a series of bottom-up stimuli (physical signals carrying the visual information and biological transduction mechanisms to the working memory towards the cognitive agencies) which are selected and set by top down signals acting as global constraints on the collective neural response which allow the recognition and the response-outputs.

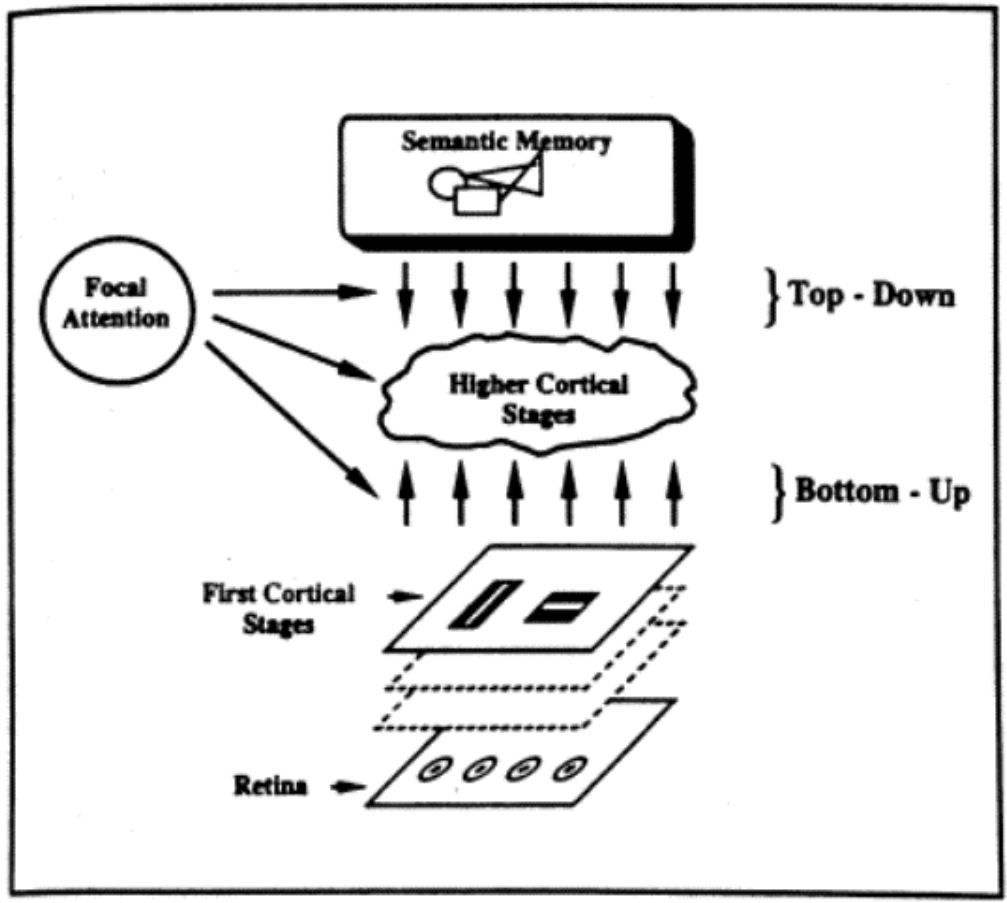

(from Julesz, 1991)

    The bottom-up activities, in principle, can be totally codified as syntactic processing according to reductionist logic, but the same does not go for the top-down features which, instead, depend on the previous memories, knowledge and aims of the observer. Without such elements there's no vision, just pattern recognition. The basic evolutionary meaning of top-down dynamic constraints - some of which are part of our deep genetic baggage – consists in filtering the stimuli related to essential information for adaptation. The top-down processes selectively amplify the expected stimuli and extinguish or soften other ones according to a priority ranking centred on the stratification of previous experiences. So, the resonance feed-back between neural agencies can be considered as an information "evaluating" so as to select a stimulus from the background noise and activate a motional and linguistic decision. On the other hand, it may happen that an impulse does not fall upon the already stored memories, in this case cognitive activity has to make the feature binding by building new categories and new interpretative codes able to create the required harmonization

    An interesting model of neural micro-dynamics for the adaptive relation between bottom-up and top-down processes uses the mechanisms of homoclinic chaos (Arecchi et al., 2002a, 2002b). Since their famous programmatic paper, the "Dynamical System Group" had put forth the hypothesis that the order/disorder peculiar mixture of non-linear dynamics could be a model for the critical sensitivity of the attentive processes regarded as information amplifying (Crutchfield et al. 1986; Licata, 2008b). Within the homoclinic chaos scenario the neuronal spikes are considered as regular orbits having erratic times. When a stimulus is recognized via an ART feed-back, such spikes get synchronized by means of a phase-linking process.

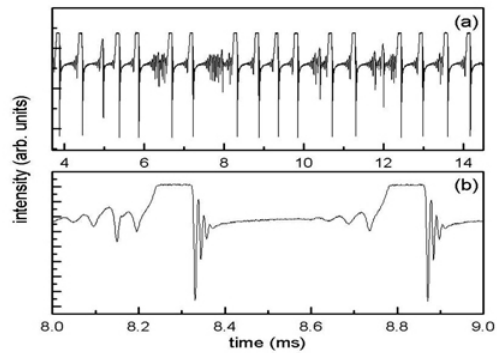

(a) Homoclinic chaos orbits with erratic times
(b) a single orbit

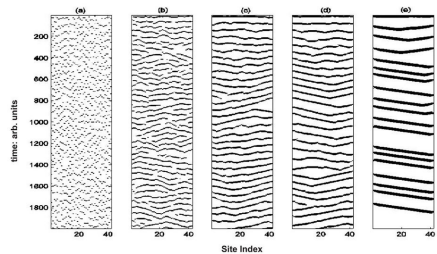

Synchronization patterns in homoclinic chaos

(from Arecchi et al., 2002a, 2002b).

The duration and the specific modalities of the coherence depend on the kind of experience and give a natural explanation for to the consciousness' time as the duration of the coherence states (Libet, 2005). S. Zeki hypothesized the activation of this kind of processes for the neural agencies related to "shape" and "motion" which come into play in the aesthetic experience (Zeki, 2000). Tangible and suggestive examples of coherence processes realized by vision are the Escher's famous lithographs or the dilemmas like "duck-rabbit" (see for ex. Jastrow, 1973). There we have a reliable performance of visual perception mechanisms, but it does not correspond to the semantic dimension of seeing; this one only occurs when coherence makes an interpretation "to collapse".

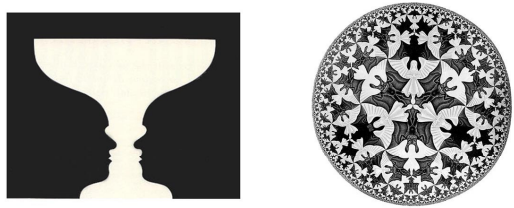

Visual cognitive dilemmas: vase or faces?   angels or devils?
(Escher's Circle Limit IV, 1960)

The D. Hofstadter "statistical mentalics" hypothesis is centred on the idea to carry out a correlation between symbolic and sub-symbolic states, like it has been done in Statistical Physics with thermodynamics and kinetic theory of gases (D. Hofstadter, 1996). Following this physical model, it has been suggested that different sub-symbolic states correspond to the same cognitive-symbolic description (Clark, 1991). One might thus be tempted to assimilate the top-down constraints with the cognitivism high level symbolic language and the visual perception bottom-up processes with the connectionism low-level one. Nevertheless, it has been showed that this kind of program can only be carried out in few, very simple cases (Smolensky, 2006), moreover the scheme here discussed provides further reasons for such hypothesis failing. Actually, the top-down constraints are not fixed schemes that can be assimilated to an algorithm, but they are neural landscape continuously and dynamically redefining by the system/environment relations. The computational descriptions of classical cognitivism (Marr, 1983) only work in "close" worlds - far from the emergence zones - when the ART process does not remodel the cognitive scenario. Some experiments on the olfactory memory of locusts and rabbits (Laurent & Wehr, 1999; Laurent et al., 1996; Walter Freeman, 2000) are quite significant. The first experiment has revealed that the temporal sequence of the neuronal activities codifying on odour does not vary when the same stimulus is presented again, whereas, in rabbits, different sequences correspond to the same stimulus; it means that the rabbit's cognitive dynamics changes at each experience so modifying its repertory of meanings. Differently from what statistical mentalics hypothesizes, *top-down and bottom-up features of vision are not two different descriptive levels, but the facets of a single dynamic process.* Therefore describing vision as a coherence process provides a way out of the classical representationalism tight corner as well as its microscopic version which is the "grandmother neuron". What we are looking at is the evolutionary dynamics of observer-centred meanings, without these ones there is just passive perception of stimuli, but not authentic vision (Arecchi, 2001; Tagliasco & Manzotti, 2008). Such analysis can be easily moved to the epistemological area of model building.

### 4. SEEING BY MODELS

There's an old tendency in the "Platonic-inspired" epistemology to be able to abstractly outlining the knowledge method, compressing it in "agnostic" form with respect to meaning and making it as much independent from the observer as possible. An explicit program of this kind has been proposed within the ambit of Neopositivism in the "Encyclopedia of unified science", which has been conceived as the empirical sciences equivalent of Whitehead and Russell "Principia Mathematica", a purely formal-logic, syntactically well defined and totally self-contained structure of science knowledge. Here too, likewise in artificial vision, the difficulties in creating an "automatic scientist, have proven to be extremely instructive about the real dynamics of science knowledge production (Thagard, 1993 ; Magnani, 2006).

Let us consider an S system studied by an Obs experimental apparatus and described by an M model. M is essentially made up of a set of variables, their evolution equations and the boundary conditions defining S. Once the $O_i$ system's observables are fixed (with i varying on a finite set), Obs can be completely specified by algorithm-like operational procedures, a Turing-Observer (see Licata, 2006). Under such conditions an M model can be regarded as an expert system manipulating the data obtained by the Turing-Observer, and the "competition" between the M models describing S can be considered a Bayesian procedure of this kind:

$$(1) \quad prob(\langle M|O\rangle) = \frac{prob\langle O|M\rangle \cdot prob\langle M|}{prob|O\rangle}$$

The formula expresses the greater capacity of one model among the others to quickly climb the probability hill's peak, so providing the maximum fitness in the space of O observables in terms of correlation and predictability according to the algorithmic compression criterion. The formula (1), applied to a single model perfectioning, can be seen as a Darwinian procedure as: *model*

*formulating – its matching to data – mutation and picking out of the maximum a posteriori probability.*

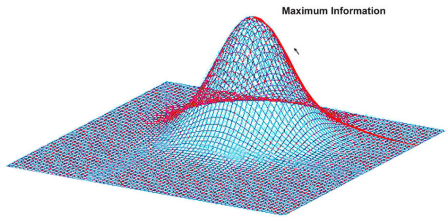

Fitness Hill for a Single Model

We can obtain a formally alternative, but conceptually similar description within the ambit of Game Theory. If we take into consideration the model with the highest score in the game whit the system's values of observables, or within Fuzzy Theory, then we can say that the winning model is the one allowing a total defuzzification of the system. Patently, if theories (or the Nature!) worked like that, natural and conceptual ecosystems would be quite poor! In actual fact, in studying phenomena, we cannot identify *a priori* the system we are dealing with and its significant variables, it depends on the model choosing, that is to say very refined and "opportunist" choices (Einstein said that the scientist must appear as a type of unscrupulous opportunist to the epistemologist!). According to the above-outlined setting, the observational-experimental context is prearranged and fixed, it "takes a picture" from a frozen perspective. The most interesting things in research happen instead when we *change the code* and choose to observe the system from different viewpoints. It means that the builder of models changes his "perspective" and the variables, and he uses a different Obs observational-experimental context. In practice, the same system can be described by a family of models, finite or countably infinite, each one "specialized" in seeing different features which mirror the possible interactions between the observer – here an active agent – and the observed system (Minati & Guberman , 2007).There the (1) defines a multi-peaked model scenario which can be regarded as an indication of the *semantic complexity* of the system.

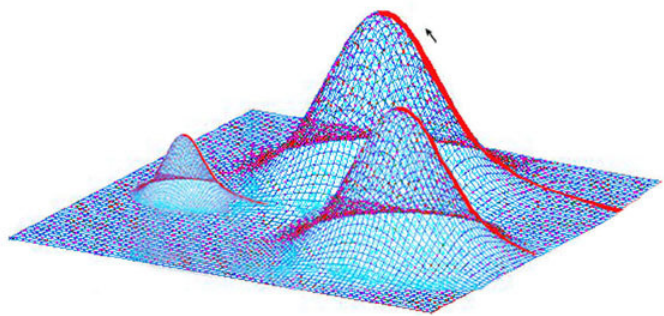

Fitness landscape for different modelling perspectives

We can write the symbolic form of the relation between a system S and the model M representing the status of the knowledge about the system by means of the Obs observer's set of choices at a given nth stage as:

$$M^{(n)} = Obs_1(S^n).$$

Obs$_1$ is an operator which guarantees the correspondence between S and M. It means that the M model "sees" the data the S system produces through the correspondence operational procedures Obs$_2$:

$$(\exp data^n) = Obs_2(M^n)$$

The acquisition of new experimental data can modify the model and require new strategies Obs$_3$:

$$M^{(n+1)} = Obs_3(\exp data^n)$$

From the above expression, we have:

$$M^{(n+1)} = Obs_1(S^{(n+1)}) = Obs_1(Obs_3(Obs_2(M^n))).$$

The operators Obs$_i$ have not to be considered as "rigid" formal tools, but as a set of model-based procedures depending on the system's nature and the observer's goals.

By putting Obs=Obs$_{1,2,3}$ and generalizing the procedure at n stages, we have the recursive formula:

$$(2) \quad M^{(n)} = Obs_n(M^{(0)}), \text{ with } n \in N.$$

The (2) sums up the analysis we have carried out: the information extraction from the system S takes place via a succession n of M models – Von Foester *eigen-models* (Von Foester,1999) – representing the "perspectives" through which the observer "watches" the system. The model is thus a cognitive filter which realizes a coherence status between the system and the observer.
Now, let us consider an infinite succession of interactions between the observer and the system:

$$M^{(\infty)} = \lim_{n \to \infty} Obs_n(M^{(0)}).$$

The natural question is: can this scenario based on models converge to a "fixed point", a unique model that somehow contains all the others? We will find out the answer - in general - is a negative one. Nevertheless, we have first to focus on the hypothesis of a positive answer, that is to say – traditionally – the reductionist one.

## 5. REDUCTIONISM'S SHORT-SIGHTEDNESS

According to the widespread reductionist approach, the theoretical scenarios considered as fundamental are those whose "arrows always point downward" towards the system's elementary constituents: particles, molecules, neurons, and so on. In spite of its great achievements, this strategy is sometimes a dead end road. If we focus our "resolution" on the elementary constituents, the information contained in the initial conditions per time unit – called Kolmogorov-Chaitin entropy – will quickly erode and the algorithmic complexity of the system will exponentially grow at the increasing of the amount of particles (Chaitin, 2007). In other words, we have to face a computational catastrophe or to change the code and to build a new model, for example based on new collective variables, in order to describe the system's patterns. When syntax grows too complicated to be tackled in detailed way, it is more useful to select significant information on different levels. This is a cognitive strategy analogous to the one taken into consideration for vision: we focus our attention on a quantity of information which is quite long-lasting to be recorded and studied.

These considerations are quite obvious in life and social-economic sciences, where the significant aspects do not merely dwell in "components", but in the functional dynamics of the structures. Most of the "interesting" phenomena we deal with need global "architectural" approaches; these ones cannot be derived from the "fundamental bricks" because systems formed by very different elements can show really similar collective behaviours whose universality is much more significant than the bare individuating of the elementary components. That is the meaning of the Anderson's famous "More is Different" (Anderson, 1979) which stressed precisely the universality of the spontaneous symmetry breaking processes in infinite state quantum systems as general conceptual frame for emergence (Pessa, 2002). Furthermore, different approaches tending to take into consideration the possibility to individuate the "constituent objects" as a consequence of the universal properties of the emergence processes have recently come out also in physics. Consequently, even the distinctions and correlations between "macroscopic state" and "microscopic state" are largely problematic and context-dependent (Licata, 2009).

The reductionism limit lies in mainly focusing on the "level" notion as well as in aiming at a world description which consists in a chain of piled-level models resembling the "Hanoi tower". Unfortunately, it is possible only within few, definite theoretical frames, such as the Effective Field Theories where the Quantum Field Theory syntax makes possible to build a chain of levels where each level individuates a scale of energy, times and lengths; it is thus possible to connect a level with the other one by proper "matching conditions" ruling the parameters of the bordering levels (Castellani, 2000).

It has to be strongly stressed that the modeling process in science is not ruled just by the notion of level; it is, above all, "goal-seeking", that is to say grasping the peculiar features of a phenomenon (Ryan, 2006). So, the above-examined problem of the "model-based" landscape can be put as follows: which are the relations between the dynamic and structural modification of a system (ontological aspect) and the model choosing (epistemic aspect)? What kind of emergence is detected by the different class of possible models? Finally, is a general theory of the observer/observed relations necessary?

## 6. BUILDING OF VISIONS: EMERGENCE AND LOGICAL OPENNESS

Building a model, as it happens for vision, realizes a cognitive homeostatic equilibrium between the observer and the observed. The type of chosen model mirrors the builder's choices about the system under examination. Let's remember we have a *logical close model* when we can always assign a value to the state variables, which means we are operating within a univocally defined syntax. That's when a computational approach to the problem can be followed and is useful.

Nevertheless, most systems we deal with are *logical open*, continuously exchange matter, energy and information with environment so reshaping their internal organization and modifying their hierarchical and functional relations (Minati, Pessa, Penna, 1998; Licata; 2008b, 2008c). A system like that cannot be solved by a closed model and is described by a *logical open* model, i.e. when there does not exist a recursive procedure to determine which information is relevant or not in describing the system behaviour. That's precisely the same as vision, where the feed-back between the bottom-up stimuli and the top-down dynamic structures leads to the emergence of new codes able to control the perception of new schemes.

Just in the same way as the Gödel Theorem shows the impossibility to compress down mathematics into axiomatic systems - mathematics is an *open system* (Chaitin, 2007) - the *Logical Openness Theory* defines the complexity degree of a system in relation to its descriptive incompressibility within a single model. Two different models of emergence correspond to logical close and open models, respectively:

- *Computational:* the formation of patterns in continuous or discrete non-linear systems, like dissipative structures ((Nicolis & Prigogine, 1989) or cellular automata (Wolfram, 2002), where the information amplifying in polynomial or exponential time can be observed. In this case, the "newness" detected by the observer is a mathematical consequence, yet not banal, of the adopted model' structure, i.e. in principle, it possible to have a local computational description of these systems, the detailed, long-term unpredictability is only linked to the critical sensibility to initial conditions and the "loss of memory" of these ones during the dynamical evolutions;

- *Intrinsic or observational:* the emerging of the system's new behaviours *cannot be predicted by the adopted model* and it requires a new formulation of the model. It is a more *radical* case than the computational one and imposes to make use of each other complementary models which focus on different features of the system depending on its behaviour and the modeler goals as well. Far from being an exotic situation, this is the norm for biological and cognitive systems. These ones show *semantic emergence* through logical openness transitions indicating the system capacity to autonomously manage information and its relation with the environment. In other words, we can say that *intrinsic emergence comes out when the very system's nature compels the observer to build new models again and again by using different cognitive strategies and dynamically managing them* ( Minati, 2008).

The logical open systems can be ordered within complexity classes depending, in general, on the thermodynamical cost their informational and physical structures meet, which has an impact on the model choosing. Without going into details, we have to remember that the more a system complexifies its structure, the more its dissipation increases; if dissipation does not destroys the system, it means that there is a set of n constraints preventing it. Within this context, the term "constraint" globally includes the significant features of the system/environment relation, such as the boundary and initial conditions, balance laws, variation of parameters and so on. Let us, thus, introduce the concept of system with n logical openness as characterized by n number of constraints, with n finite. We can draw it as a graph with n vertices, each representing, from thermodynamical viewpoint, an entropy containment mechanism and, from informational viewpoint, a specific informational path by which the system processes the inputs in outputs.

It is easy to give a formal demonstration that a) *it is impossible to describe a n logical open system by a single model* and b) *describing an n degree open system of by a model with m logical openness, where m<n has always a limited validity domain*. We are going to focus here on the likeness between these results and the formal logic limiting ones and, putting aside the mathematical details, to concentrate on their conceptual meaning by taking up again the graph image. Adopting a single model means to fix variables and interactions, i.e. a finite and fix number of n vertices, whereas in a highly logical open system they continuously and unpredictably emerge and disappear as a consequence of the *internal* functional organization of the system. So, adopting a model is an arbitrary partition the observer does on the system/environment relation.

The *model-building activity itself, like any cognitive activity, is an open system* which cannot be described within the syntax of a "unique" model, but by means of a plurality of co-adaptive processes between the observer and the observed (Maturana & Varela, 1992). A "theory of everything" is impossible for this kind of systems. It would mean to hypothesize an infinite logical openness, i.e. the existence of a sort of Laplacian super-observer able to describe each state at each instant of the system/world relation; it is nothing else but the reductionist utopia. We get no the God's Eye! (Vamos, 1991) Such idea is in consonance with the interesting Breuer theorem; he states that no theory, both classical or quantum, can describe each state of a system where the observer is excluded (Breuer 1995, 1997).

Reductionism is a good strategy for those systems whose resolution fits models with low logical openness, within a one-to-one correspondence between syntax and semantics. If we now look at cognitive models, we will find that AI hints at an observer using logical closure where the number of constraints is low, time independent and thus producing inaccessible or "opaque" information (Clark, 1991). Connectionist models, instead, are placed at a higher level of logical openness; the learning of a supervised neural net, in fact, only slightly depends on the initial "genetic program" and it gets complexified in interacting with the environment. The ultimate level is "quantum brain", a model with very high logical openness which takes into account the emerging of new codes and semantic domains by means of the dissipative relations with the environment (Vitiello, 2001).

## 7. SCIENCE IN THE TIME OF COMPLEXITY

Both vision and building world-models are features of adaptation processes implying the involvement of a cognitive filter, the activation of a semantic space which makes the representation itself possible. We have not to intend it as a "photograph" of the world, but as a dynamic game between the eye and the world occurring by high logical open, creative strategies. The observer is part of the description by its making choices like Velázquez is inside his famous *Las Meninas*. This semantic complexity reflects the possible infinite states of homeo-cognitive equilibrium between the observer and the observed. What does all that tell us about science?

The received view of classical epistemology has always had a normative and linear character which has never seriously thrown into crisis – except for "local" one - the possibility to provide an objective world representation achieved by the deep stratification of the theoretical fabric as well as the Darwinian selection of the "right" models able to pass the experience test (saving the phenomena) and to connect with the "fundamental" theories, such as Relativity and Quantum Mechanics in Physics and Natural Selection and the central dogma of molecular biology in Life Sciences. Occasionally, a single, problem-stemming model could undermine the theoretical panorama at its bottom and modifying it, such as the Planck hypothesis on the black-body radiation, anyway the idea of an asymptotic approximation towards a unified and definitive theory of the world remained untouched. It has caused not only the drastic splitting between hard and soft sciences - so ratifying the Cartesian divorce between mind sciences and matter sciences – but it has endorsed a simplified idea of the relations between knowledge processes and the world, the former regarded as method and the latter as a "code" to crack.

On the other hand, some extreme forms of radical constructivism shift toward the relativistic-flavoured idea of eliminating the object. A simplified idea, as well. In fact, if we decide that a model works, we are stating something as 'a key opens a keyhole' and, above all, that another model or another key do not work! Stating that the observer creates all that means to recognize the context-dependent nature of the observer/observed relation. If so, the notion of "producing a world" must include a unitary vision of the *adaptive relations* between the observer and the observed, rather than a dangerous ontological leaning toward either the object or the subject. The problem with radical constructivism is that it does not seem to find authentic explanations for the unitary dynamics of science, that is to say the capacity of our models to cluster into structures and meta-structures, in syntactic classes and theoretical chains: there actually exist keys which open more than a door! ( See for ex.: Coniglione, 2008).

Generally, the term "ontology" gives scientists hives since they are especially interested in building the conceptual and formal tools able to identify the problems and try to answer them. The viewpoint of traditional epistemologies is coarse-grained. In everyday reality of research, instead, many tendencies and micro-paradigms compete for the "saving of phenomena" and – except for few really well-defined syntaxes, such as Quantum Field Theory and the Standard Model – the so-called fundamental and most radicated theories, previous memories of scientific knowledge, do not help in univocally selecting among models, whose differences hardly lie in the level or comprehensiveness of the explanation they provide, but in their goals.

Not by chance, D. Deutsch refers to fundamental theories as the "fabric of reality", a set of leading-principles acting on emergent models as general boundary conditions (Deutsch, 1998). In addition, it is worthy to remember that the new acquisitions often do not modify the "form" of fundamental theories, but our interpretation, our way to use them in order to build new knowledge.

The image of science is not that of a continent, but an archipelago, where we can see big islands, the most ancient and syntactically defined ones, and – at the same - the emerging of smaller islands, maybe transient, where the conceptual bridges between different islands are built again and again.

To fulfil the requirements of complexity sciences, the scientific activity has to be able to comprehend the adaptive dynamics between system and environment as well as the model and the context where it is applied. Adaptive epistemology based on logical openness overcomes the thigh corners of both naïve objectivist conception and radical relativism temptations by stressing that the knowledge building is a process of continuous shifting from the "frozen" syntactical dimensions to the plurality choices which make possible for mind and world to meet each other.

## AKNOWLEDGEMENTS
The author thanks G. Minati and T. Arecchi for the precious conversations throughout the years about these topics. This paper is dedicated to Vincenzo Tagliasco (1941 – 2008)